\documentclass[aps,prb,twocolumn,10pt,showkeys,superscriptaddress]{revtex4-2}
\usepackage[T1]{fontenc}
\usepackage[utf8]{inputenc}
\usepackage{microtype} 
\usepackage{newtxtext,newtxmath}
\usepackage{physics}   
\usepackage{graphicx}
\usepackage[dvipsnames]{xcolor}
\usepackage{booktabs}
\usepackage{multirow}
\usepackage{orcidlink}
\usepackage[normalem]{ulem}
\usepackage{cancel}
\usepackage{appendix}
\usepackage[dvipsnames]{xcolor}

\usepackage{hyperref}
\usepackage[table]{xcolor}
\definecolor{myMagenta}{RGB}{220,0,110}  
\hypersetup{
  colorlinks=true,
  linkcolor=blue, citecolor=blue,
  filecolor=blue, urlcolor=blue,
  breaklinks=true
}

\begin{document}
\title{Thermal Analysis, Joule-Thomson Expansion and Hawking Sparsity of Mod(A)Max-AdS Black Hole Immersed in a Cloud of Strings}

\author{Faizuddin Ahmed\orcidlink{0000-0003-2196-9622}}
\email{E-mail: faizuddinahmed15@gmail.com}
\affiliation{Department of Physics, The Assam Royal Global University, Guwahati, 781035, Assam, India}

\author{Ahmad Al-Badawi\orcidlink{0000-0002-3127-3453}}
\email{E-mail: ahmadbadawi@ahu.edu.jo}
\affiliation{Department of Physics, Al-Hussein Bin Talal University, 71111, Ma'an, Jordan}

\author{Edilberto O. Silva\orcidlink{0000-0002-0297-5747}}
\email{E-mail: edilberto.silva@ufma.br}
\affiliation{Programa de Pós-Graduação em Física \& Coordenação do Curso de Física-Bacharelado,
Universidade Federal do Maranhão, 65085-580 São Luís, Maranhão, Brazil}

\date{\today}

\begin{abstract}
We investigate the thermodynamic behavior of a spherically symmetric Anti-de Sitter black hole in Mod(A)Max electrodynamics surrounded by a cloud of strings. Within the extended phase-space framework, we treat the cosmological constant as a pressure and interpret the black-hole mass as enthalpy, which enables a unified discussion of local stability, global phase structure, and Joule--Thomson expansion. We analyze the Hawking temperature, Gibbs free energy, and heat capacity, and show how the string-cloud parameter, the Mod(A)Max deformation, and the electric charge reshape the physical domain, the stability windows, and the small/large black-hole transition pattern. We further characterize the critical behavior and demonstrate that a van der Waals--like phase structure arises only in the physical sector, while the alternate branch does not admit a genuine critical point. For the Joule-Thomson process, we determine the inversion curve and the corresponding isenthalpic trajectories, highlighting how the model parameters control the cooling/heating regimes and can generate terminating isenthalpic behavior at sufficiently large charge. Finally, we examine the sparsity of Hawking radiation and discuss how the underlying parameters influence the temporal discreteness of the emitted flux, particularly near extremality and in the large-radius AdS regime.
\end{abstract}

\maketitle

\section{Introduction}

Black-hole thermodynamics has become one of the most fertile arenas where gravity, quantum theory, and statistical physics meet. Since the discovery that black holes emit thermal radiation \cite{Hawking1975} and possess an entropy proportional to the horizon area \cite{Bekenstein1973}, it has been clear that horizons behave as genuine thermodynamic systems whose macroscopic variables are rooted in microscopic degrees of freedom that remain partially unknown. In asymptotically Anti-de Sitter (AdS) space-times, the thermodynamic description becomes even richer because AdS black holes can reach stable equilibrium with a heat bath and exhibit robust phase structure, including transitions that resemble ordinary condensed-matter phenomena. A particularly influential development is the extended phase space (or black-hole chemistry) program, in which the cosmological constant is promoted to a thermodynamic pressure and the black-hole mass is reinterpreted as enthalpy \cite{DK2009,KubiznakMann2012}. This perspective not only clarifies the role of the thermodynamic volume, but also exposes Van der Waals-like criticality and universal scaling, enabling systematic comparisons between gravitational systems and standard fluids.

In parallel with the gravitational side, nonlinear electrodynamics (NLED) has long been recognized as a physically motivated deformation of Maxwell theory, originally introduced to tame field singularities and to model strong-field regimes. In the context of black holes, NLED deformations are known to significantly alter thermodynamic response functions, phase structure, and stability boundaries, and they provide a controlled way to parameterize departures from the linear electromagnetic sector. Among the modern NLED models, ModMax (and its extensions) occupies a special place because it preserves conformal invariance and electric-magnetic duality in four dimensions while remaining genuinely nonlinear \cite{IB2020,BPK2020}. 
These symmetry properties make Mod(A)Max electrodynamics an attractive arena for exploring strong-field effects without sacrificing key structural features of Maxwell theory. When coupled to AdS gravity, Mod(A)Max deformations can modify the effective charge sector and thereby reshape the competition among the geometric term, the AdS pressure term, and the electromagnetic contribution in the Hawking temperature and related thermodynamic potentials. As a consequence, one expects quantitative and, in some parameter domains, qualitative changes in local stability (heat capacities), global phase preference (Gibbs free energy), and critical properties in the extended phase space.

Another physically motivated ingredient that has received renewed attention is the inclusion of string-like matter sources. A convenient effective description is provided by a cloud of strings, which modifies the geometry by inducing a solid angle deficit and contributes an additional parameter controlling the horizon structure and thermodynamic behavior \cite{PSL1979}. From a phenomenological standpoint, such sources serve as proxies for networks of cosmic strings or stringy remnants in the environment of compact objects, and they offer an analytically tractable way to investigate how external matter distributions deform the black-hole equation of state. In the extended phase space, these deformations are particularly interesting because they can shift the location of critical points, change the coexistence window of small/large black-hole phases, and affect the onset (or suppression) of first-order transitions.

Beyond equilibrium thermodynamics, two directions have become especially relevant for building a more complete physical picture. The first is the study of thermodynamic engines built from black-hole working substances. In black-hole chemistry, one can define cyclic processes in the $P$-$V$ plane and compute efficiencies, thereby treating AdS black holes as heat engines. This viewpoint has led to a broad set of results connecting gravitational phase structure, response functions, and engine performance \cite{Johnson2014}. The second direction is the analysis of non-equilibrium expansion processes, among which the Joule-Thomson (JT) expansion has emerged as a standard diagnostic of cooling/heating behavior in AdS black holes. In ordinary fluids, JT expansion separates cooling and heating regions via an inversion curve; in black holes, the corresponding inversion curve in the $T$-$P$ plane is sensitive to charge, rotation, and model deformations. The earliest systematic discussions of JT expansion for charged AdS black holes and subsequent generalizations established that the inversion structure provides a refined probe of the interplay between the equation of state and enthalpy constraints in extended thermodynamics \cite{OO2017,SQL2018,CL2020}. Therefore, investigating JT expansion in Mod(A)Max backgrounds with additional matter sources (such as a cloud of strings) is both timely and well-motivated.

A further aspect that has gained prominence is the role of thermal fluctuations and the resulting corrections to black-hole entropy and stability criteria. Even when the classical Bekenstein-Hawking entropy provides the leading contribution, statistical fluctuations around equilibrium generically induce logarithmic corrections, which can become important for small black holes or near critical regimes where susceptibilities grow \cite{DasMajumdarBhaduri2002}. From the perspective of phase structure, such corrections may shift stability boundaries, modify response functions, and alter the effective critical behavior, thus providing a complementary window into the robustness of equilibrium predictions under subleading statistical effects.

Finally, Hawking radiation-despite being thermal in spectrum-is intrinsically discrete as an emission process: quanta are produced sporadically rather than as a continuous flux. This motivates the notion of Hawking sparsity, a dimensionless diagnostic comparing the characteristic thermal wavelength to the effective emitting area, thereby quantifying how dilute the emission is in time \cite{Page1976df,Gray2016}. In deformed gravity or matter sectors, sparsity becomes an incisive probe because it depends nonlinearly on the temperature and horizon geometry; thus, it can amplify differences between models that may appear modest at the level of, e.g., $T(r_h)$ alone. In AdS settings, where large black holes can be thermodynamically stable, sparsity analysis complements standard thermodynamic quantities by highlighting how emission characteristics reorganize across parameter space.

Motivated by these developments, in this work we investigate the thermodynamic behavior of a spherically symmetric AdS black hole sourced by Mod(A)Max electrodynamics and surrounded by a cloud of strings. We analyze the equilibrium sector through the Hawking temperature, Gibbs free energy, and heat capacity, emphasizing how the cloud-of-strings parameter and the Mod(A)Max nonlinearity deform local stability and global phase preference. 
Within the extended phase space, we explore the associated criticality and the parameter dependence of the critical point, clarifying the regimes in which Van der Waals-type behavior is realized. We then study Joule--Thomson expansion via inversion curves and isenthalpic trajectories, using the mass/enthalpy identification to map the cooling/heating boundary and its sensitivity to the charge and Mod(A)Max parameters. We also connect these results with heat-engine considerations in AdS black-hole chemistry, discussing how the deformed thermodynamics constrains efficiency trends. 
Finally, we quantify the sparsity of Hawking radiation and examine how model parameters control the degree of temporal discreteness of emission. 
Altogether, our analysis provides a coherent thermal portrait of Mod(A)Max-AdS black holes in a string-cloud environment, offering multiple complementary diagnostics (criticality, JT expansion, engine behavior, and sparsity) for characterizing the interplay between nonlinear electrodynamics and external string-like matter sources.

\section{Spherically Symmetric Mod(A)Max-AdS BH with a Cloud of Strings}

In this paper, we study a spherically symmetric Mod(A)Max black hole coupled with a cloud of strings. The action describing the coupling of Einstein gravity with the Mod(A)Max electrodynamic fields and matter field (string clouds) is given by
\begin{equation}
\mathcal{S} = \frac{1}{16\pi} \int_{\mathcal{M}} d^4x \, \sqrt{-g} \left(R - 4 \eta \, \mathcal{L} \right) + \mathcal{S}_{\rm CS},\label{a1}
\end{equation}
where $g = \det(g_{\mu\nu})$ is the determinant of the metric tensor $g_{\mu\nu}$, $R$ is the Ricci scalar, and $\mathcal{S}_{\rm CS}$ represents the action matter fields and we consider a cloud of strings as field in our investigation.

In the action above, $\mathcal{L}$ denotes the ModMax Lagrangian, defined as \cite{IB2020,BPK2020}
\begin{equation}
\mathcal{L} = \mathcal{S} \cosh \gamma - \sqrt{\mathcal{S}^2 + \mathcal{P}^2} \, \sinh \gamma,\label{a2}
\end{equation}
where $\gamma$ is the dimensionless ModMax parameter. The quantities $\mathcal{S}$ and $\mathcal{P}$ are the electromagnetic scalar and pseudoscalar invariants, respectively, defined by
\begin{equation}
\mathcal{S} = \frac{\mathcal{F}}{4}, \qquad \mathcal{P} = \frac{\widetilde{\mathcal{F}}}{4},\label{a3}
\end{equation}
with $\mathcal{F} = F_{\mu\nu} F^{\mu\nu}$ and $\widetilde{\mathcal{F}} = F_{\mu\nu} \widetilde{F}^{\mu\nu}$. Here, $F_{\mu\nu} = \partial_\mu A_\nu - \partial_\nu A_\mu$ is the electromagnetic field tensor, $A_\mu$ is the gauge potential, and $\widetilde{F}^{\mu\nu} = \frac{1}{2} \epsilon^{\mu\nu\rho\lambda} F_{\rho\lambda}$ is its dual. By setting $\gamma = 0$, the Lagrangian reduces to the standard Maxwell form: $\mathcal{L} = \mathcal{F}/4$. 

To construct electrically charged black hole solutions, we consider $\mathcal{P} = 0$. Furthermore, we include a cloud of strings contribution as the matter field. Consequently, the generalized Einstein field equations in the presence of Mod(A)Max electrodynamics and CS read \cite{DFA2021,PSL1979}:
\begin{equation}
G_{\mu\nu} = 8\pi T_{\mu\nu} + T^{\rm CS}_{\mu\nu},\label{a4}
\end{equation}
where $T_{\mu\nu}$ is the energy-momentum tensor associated with the Mod(A)Max field, and $T^{\rm CS}_{\mu\nu}$ corresponds to CS. The Maxwell equations for the charged case take the form
\begin{equation}
\partial_\mu \left( \sqrt{-g} \, e^{-\gamma} F^{\mu\nu} \right) = 0.\label{a5}
\end{equation}

To include string-like objects, we consider Nambu-Goto action given by \cite{PSL1979}
\begin{equation}
    S_{\rm CoS}=\int \sqrt{-\gamma}\,\mathcal{M}\,d\lambda^0\,d\lambda^1=\int \mathcal{M}\sqrt{-\frac{1}{2}\,\Sigma^{\mu \nu}\,\Sigma_{\mu\nu}}\,d\lambda^0\,d\lambda^1,\label{a6}
\end{equation}
where $\mathcal{M}$ is the dimensionless constant which characterizes the string, ($\lambda^0\,\lambda^1$) are the time
like and spacelike coordinate parameters, respectively \cite{JLS1960}. $\gamma$ is the determinant of the induced metric of the strings world sheet given by $\gamma=g^{\mu\nu}\frac{ \partial x^\mu}{\partial \lambda^a}\frac{ \partial x^\nu}{\partial \lambda^b}$.  $\Sigma_{\mu\nu}=\epsilon^{ab}\frac{ \partial x^\mu}{\partial \lambda^a}\frac{ \partial x^\nu}{\partial \lambda^b}$ is bivector related to string world sheet, where $\epsilon^{ ab}$ is the second rank Levi-Civita tensor which takes the non-zero values as $\epsilon^{ 01} = -\epsilon^{ 10} = 1$.

The energy-momentum tensor of string-like objects is given by
\begin{equation}
   T_{\mu\nu}^{\rm CS}=2 \frac{\partial}{\partial g_{\mu \nu}}\mathcal{M}\sqrt{-\frac{1}{2}\Sigma^{\mu \nu}\,\Sigma_{\mu\nu}} =\frac{\rho^{\rm CoS} \,\Sigma_{\alpha\nu}\, \,\Sigma_{\mu}^\alpha }{\sqrt{-\gamma}}, \label{a7}
 \end{equation}
where $\rho^{\rm CS}$ is the proper density of the string cloud. The energy-momentum tensor components are given by
\begin{equation}
    T^{t\,(\rm CS)}_{t}=\rho^{\rm CS}=\frac{\alpha}{r^2}=T^{r\,(\rm CS)}_{r},\quad T^{\theta\,(\rm CS)}_{\theta}=T^{\phi\,(\rm CS)}_{\phi}= 0,\label{a8}
\end{equation}
$\alpha$ is the string cloud parameter.

Thereby, incorporating cloud of strings as matter field, a spherically symmetric Mod(A)Max-AdS black hole is described by the following line element \cite{BES2026,PSL1979}
\begin{equation}
    ds^2 = -f(r)\,dt^2 + \dfrac{dr^2}{f(r)} + r^2 (d\theta ^2 + \sin ^2{\theta }\,d\varphi ^2),\label{metric}
\end{equation}
where the lapse function is given by
\begin{align}
     f(r) = 1-\alpha-\frac{2 M}{r}+\eta\,\frac{e^{-\gamma}\,Q^2}{r^2}-\frac{\Lambda}{3}\,r^2,\label{function}
\end{align}
Here $M$ represent mass of the black hole and $\eta=\pm\,1$ (plus sign corresponds to ModMax and minus for phantom ModMax or Mod(A)Max black hole). It is worth noting that the presence of string-like objects produces a solid angle deficit analogue to the global monopole geometry \cite{MB1989}.

In the limit $\eta=+1$ and $\alpha=0$, corresponding to the absence of string-like objects, the function simplifies as
\begin{align}
     f(r) = 1-\frac{2 M}{r}+\frac{e^{-\gamma}\,Q^2}{r^2}-\frac{\Lambda}{3}\,r^2.\label{function2}
\end{align}
In that limit, the metric (\ref{metric}) reduces to the ModMax-AdS black hole solution whose thermodynamic properties were studied in \cite{SNG2025,GB2025} and thermal properties with a perfect fluid dark matter in \cite{YS2025,SS2025}.  

\section{Thermodynamics}

In this section, we study thermodynamic properties of the Mod(A)Max-AdS black hole coupled with CS and derive key quantities, such as th Hawking temperature, Gibbs free energy and specific heat capacity.

The black hole mass can be obtained using the condition $f(r_h)=0$, where $r_h$ is the event horizon. Using the metric function in (\ref{function}), we find
\begin{equation}
    M=\frac{r_h}{2}\left(1-\alpha+\eta\,e^{-\gamma}\,\frac{Q^2}{r^2_h}-\frac{\Lambda}{3}\,r_h^2\right).\label{b1}
\end{equation}

In extended phase space, the cosmological constant $\Lambda$ is considered as thermodynamic pressure $P$ and they are related by
\begin{equation}
    \Lambda=-8\pi P.\label{b2}
\end{equation}
Therefore, the black hole considering an internal energy of the thermodynamic system can be re-written as,
\begin{equation}
    M=\frac{r_h}{2}\left(1-\alpha+\eta\,e^{-\gamma}\,\frac{Q^2}{r^2_h}+\frac{8\pi P}{3}\,r_h^2\right).\label{b3}
\end{equation}

The Hawking temperature for a spherically symmetric black hole is given by
\begin{equation}
    T=\frac{f'(r_h)}{4\pi}=\frac{1}{4\pi r_h}\left[1-\alpha-\eta\,e^{-\gamma}\frac{Q^2}{r_h^2}+8\pi P r_h^2\right].\label{b4}
\end{equation}

\begin{figure*}[tbhp]
    \centering
    \includegraphics[scale=0.45]{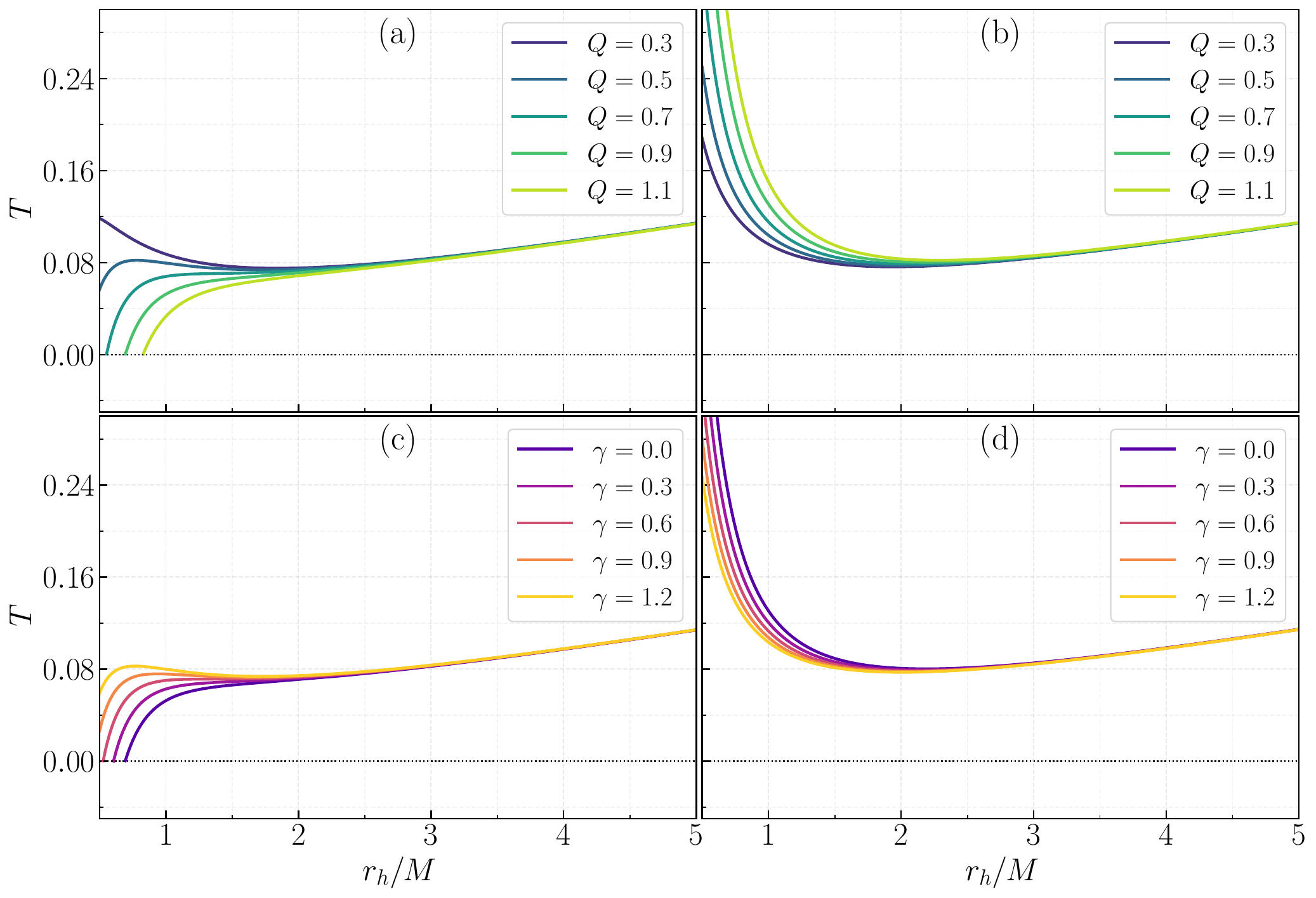}
    \caption{Hawking temperature as a function of the horizon radius for the Mod(A)Max--AdS black hole surrounded by a cloud of strings. In panels (a) and (b) we plot $T(r_h)$ for fixed $(\gamma,\alpha,P)=(0.5,0.1,0.01)$ and several charge values $Q=\{0.3,0.5,0.7,0.9,1.1\}$, for $\eta=+1$ and $\eta=-1$, respectively. Panels (c) and (d) display $T(r_h)$ for fixed $(Q,\alpha,P)=(0.7,0.1,0.01)$ and $\gamma=\{0.0,0.3,0.6,0.9,1.2\}$, again for $\eta=+1$ and $\eta=-1$, respectively. The zeros of $T$ separate physical ($T>0$) and nonphysical ($T<0$) branches and determine the extremal limit.}
    \label{fig:hawking_temperature_4panels}
\end{figure*}
The Hawking temperature provides the basic local thermodynamic diagnostic for the Mod(A)Max--AdS black hole in the presence of a string cloud. Figure~\ref{fig:hawking_temperature_4panels} displays $T(r_h)$ for both sectors $\eta=\pm1$, highlighting how the parameters $(Q,\gamma,\alpha,P)$ control the existence of physical branches and the extremal limit. In each panel, the region with $T>0$ corresponds to thermodynamically admissible configurations, whereas $T<0$ is discarded. Hence, the zeros of $T$ define the extremal horizon radius $r_h^{\rm (ext)}$ that separates nonphysical and physical branches.

Panels (a) and (b) show that increasing the charge parameter $Q$ enhances the magnitude of the electromagnetic contribution and shifts the temperature profile, typically enlarging the low--$r_h$ region where the temperature becomes negative and thereby pushing $r_h^{\rm (ext)}$ to larger values. This behavior is consistent with the standard competition between the charge term, which lowers $T$ at small radii, and the AdS pressure term, which raises $T$ for sufficiently large $r_h$ through the $8\pi P\,r_h^{2}$ contribution. Panels (c) and (d) illustrate the effect of the dilatonic-like parameter $\gamma$: increasing $\gamma$ suppresses the effective charge contribution via the factor $e^{-\gamma}$, which shifts the curves upward at small $r_h$ and tends to reduce the size of the negative-temperature region, consequently lowering $r_h^{\rm (ext)}$ for fixed $(Q,\alpha,P)$. Comparing the $\eta=+1$ and $\eta=-1$ sectors, one finds a quantitative change in the balance between these terms, which is reflected in distinct temperature profiles and extremality conditions. Overall, Fig.~\ref{fig:hawking_temperature_4panels} delineates the physically allowed domain in parameter space and sets the stage for the phase-structure analysis based on $G(T)$ and response functions.

The entropy of the thermodynamic system is given by
\begin{equation}
    S=\int \frac{dM}{T}=\int 2\pi r_h dr_h=\pi r^2_h=\mathcal{A}/4\label{b5}
\end{equation}
which is a quarter of the horizon area $\mathcal{A}$, satisfying the Bekestein-Hawking entropy law.

The Gibbs free energy of the system by considering the black hole mass as internal energy is given by
\begin{equation}
    G=M-T\,S=\frac{r_h}{4}\left[1-\alpha+\frac{3\eta e^{-\gamma}Q^2}{r_h^2}
-\frac{8\pi P}{3}r_h^2\right].\label{b6}
\end{equation}
\begin{figure*}[tbhp]
    \centering
    \includegraphics[scale=0.45]{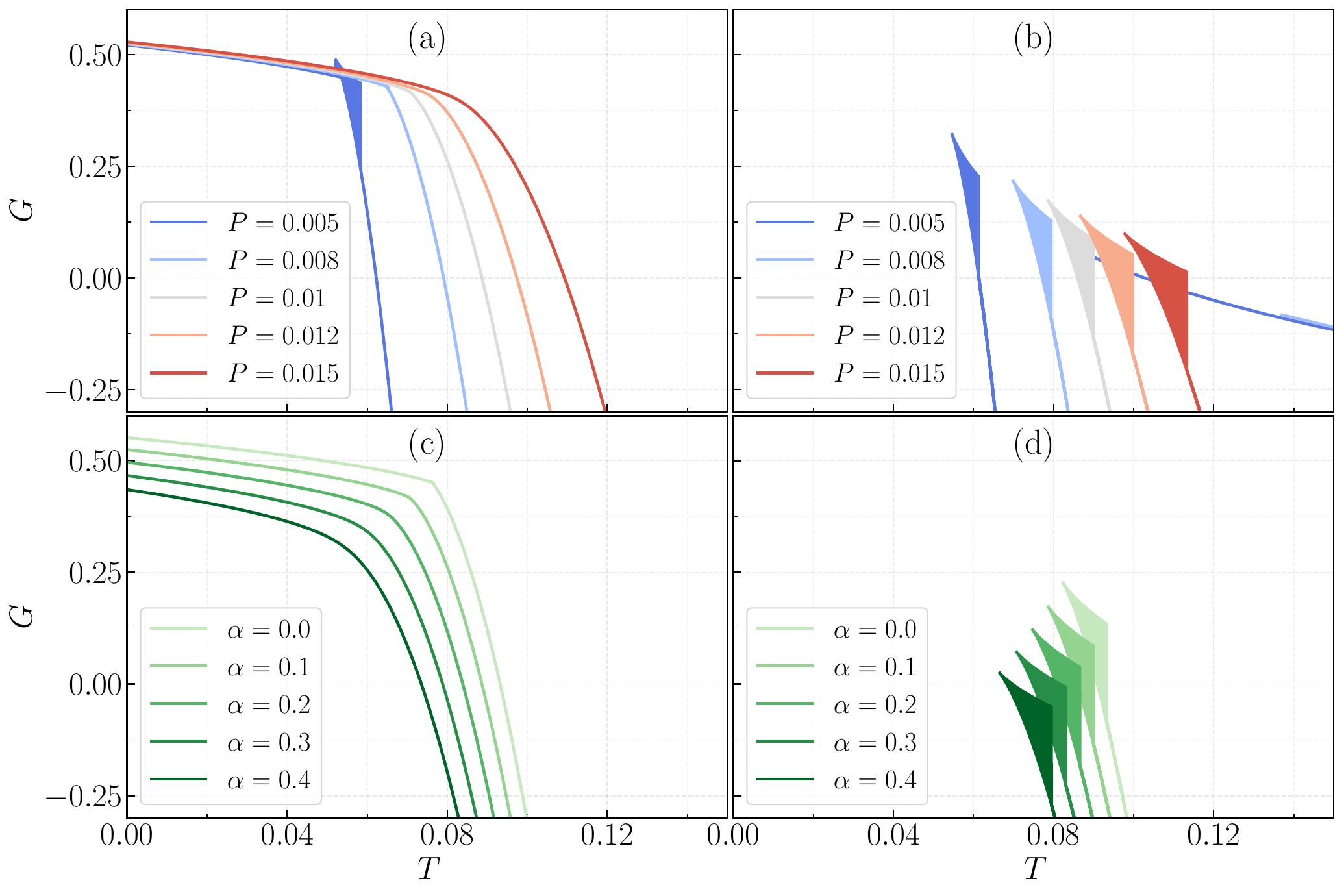}
    \caption{Gibbs free energy as a function of the Hawking temperature for the Mod(A)Max--AdS black hole surrounded by a cloud of strings. Panels (a) and (b) display $G(T)$ for fixed $(Q,\gamma,\alpha)=(0.7,0.5,0.1)$ and several pressures $P=\{0.005,0.008,0.010,0.012,0.015\}$, for $\eta=+1$ and $\eta=-1$, respectively. Panels (c) and (d) show $G(T)$ for fixed $(Q,\gamma,P)=(0.7,0.5,0.01)$ and distinct cloud-of-strings parameters $\alpha=\{0.0,0.1,0.2,0.3,0.4\}$, again for $\eta=+1$ and $\eta=-1$, respectively. The swallowtail structure (when present) signals a first-order small/large black-hole phase transition, while its disappearance indicates the approach to the critical behavior.}
    \label{fig:gibbs_4panels}
\end{figure*}
The thermodynamic phase structure can be conveniently diagnosed from the Gibbs free energy $G(T)$ in the extended phase space. Figure~\ref{fig:gibbs_4panels} shows $G$ as a function of the Hawking temperature for representative values of the pressure and of the cloud-of-strings parameter. For $\eta=+1$, the curves may develop a characteristic swallowtail profile in a certain pressure range, which is the hallmark of a first-order transition between small and large black-hole branches at a given pressure. In this case, the physically preferred phase corresponds to the global minimum of $G$ at fixed $(T,P)$, and the transition temperature is determined by the intersection of the competing branches (equal Gibbs free energy). As the pressure is increased toward the critical regime, the swallowtail shrinks and eventually disappears, indicating the termination of the first-order line at the critical point, beyond which the system displays a smooth crossover between branches.

The cloud-of-strings parameter $\alpha$ also shifts the $G(T)$ profile by modifying the competition between branches. In particular, varying $\alpha$ at fixed $(Q,\gamma,P)$ changes the temperature window where multiple branches coexist and hence may move the transition temperature and alter the prominence of the swallowtail. The $\eta=-1$ sector, shown in Figs. \ref{fig:gibbs_4panels}(b) and \ref{fig:gibbs_4panels}(d), exhibits a qualitatively distinct behavior: depending on the parameters, the swallowtail structure can be suppressed or significantly weakened, signaling that the corresponding branch structure and/or stability conditions differ from the $\eta=+1$ case. Overall, Fig.~\ref{fig:gibbs_4panels} summarizes how the extended thermodynamic variables and the string-cloud parameter control the first-order transition and the emergence (or absence) of criticality in the Mod(A)Max--AdS black hole.

Finally the specific heat capacity at constant pressure is given by
\begin{align}
    C_P=\frac{dM}{dT}=-2\pi r^2_h\,\left(\frac{1-\alpha-\eta e^{-\gamma}\dfrac{Q^2}{r_h^2}+8\pi P r_h^2}{
1-\alpha-3\eta e^{-\gamma}\dfrac{Q^2}{r_h^2}-8\pi r^2_h P}\right).\label{b7}
\end{align}
\begin{figure*}[tbhp]
    \centering
    \includegraphics[scale=0.45]{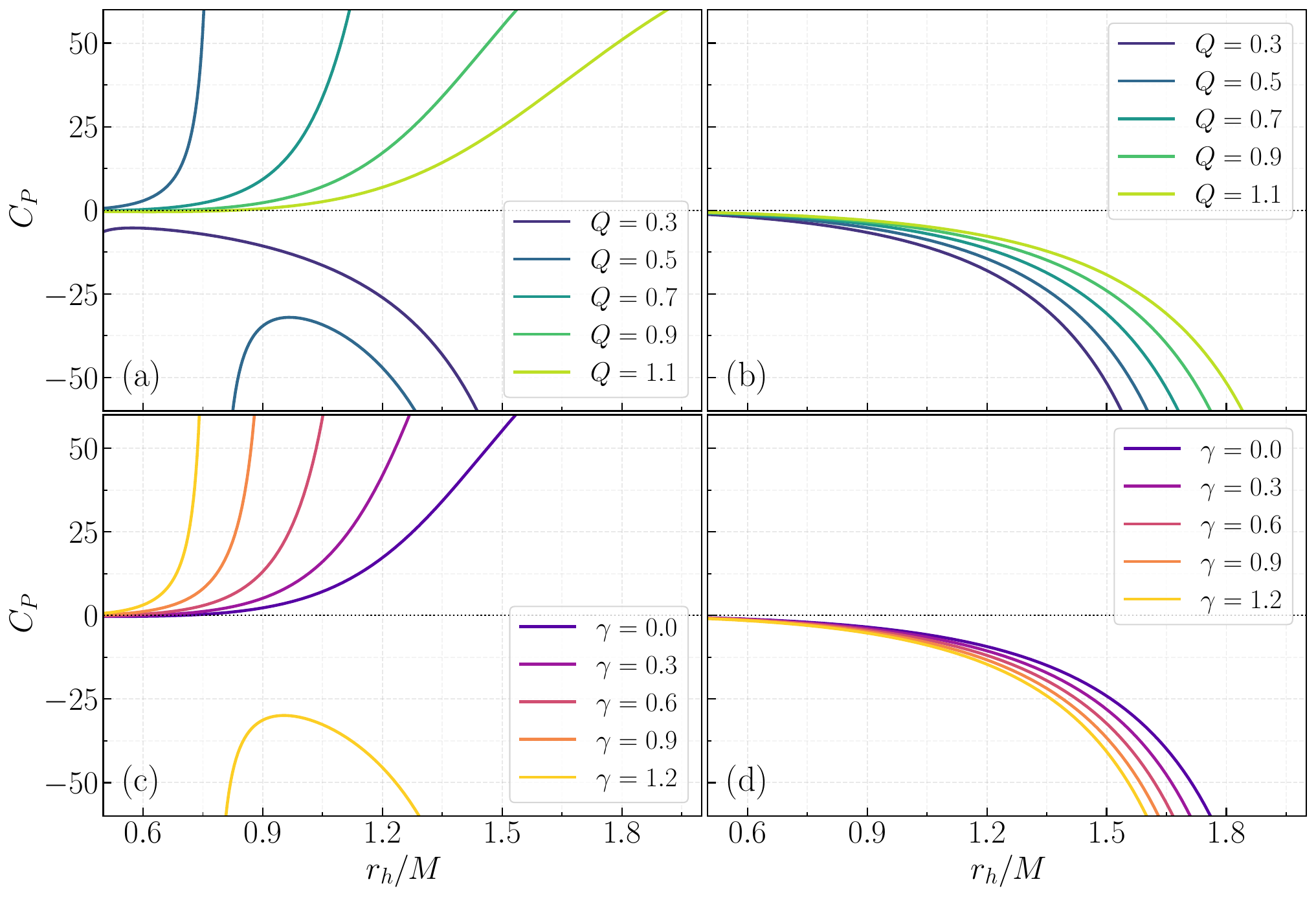}
    \caption{Heat capacity at constant pressure as a function of the horizon radius for the Mod(A)Max--AdS black hole surrounded by a cloud of strings. Panels (a) and (b) show $C_{P}(r_h)$ for fixed $(\gamma,\alpha,P)=(0.5,0.1,0.01)$ and several charge values $Q=\{0.3,0.5,0.7,0.9,1.1\}$, for $\eta=+1$ and $\eta=-1$, respectively. Panels (c) and (d) display $C_{P}(r_h)$ for fixed $(Q,\alpha,P)=(0.7,0.1,0.01)$ and $\gamma=\{0.0,0.3,0.6,0.9,1.2\}$, again for $\eta=+1$ and $\eta=-1$, respectively. The sign of $C_{P}$ distinguishes locally stable ($C_{P}>0$) from unstable ($C_{P}<0$) branches, while divergences indicate second-order transition points associated with changes in local stability.}
    \label{fig:specific_heat_4panels}
\end{figure*}
Local thermodynamic stability at fixed pressure is governed by the heat capacity $C_{P}$, whose sign determines whether the black-hole branch is stable against small thermal fluctuations. Figure~\ref{fig:specific_heat_4panels} depicts $C_{P}(r_h)$ for representative parameter choices in the $\eta=\pm1$ sectors. In all panels, regions with $C_{P}>0$ correspond to locally stable configurations, whereas $C_{P}<0$ identifies unstable branches. Moreover, the vertical divergences of $C_{P}$ signal points where the denominator of Eq.~(17) vanishes, i.e., where $(\partial T/\partial r_h)_{P}=0$; such singularities mark the onset of a change in local stability and are associated with second-order (critical) behavior in the response functions.

Figures \ref{fig:specific_heat_4panels}(a) and \ref{fig:specific_heat_4panels}(b) illustrate how the charge parameter $Q$ reshapes the stability window. Increasing $Q$ strengthens the effective electromagnetic contribution and shifts the location of the $C_{P}$ divergences, thereby modifying the range of horizon radii for which the black hole is locally stable. In particular, depending on $(Q,\gamma,\alpha,P)$, the system may exhibit a small--black-hole stable branch separated from a large--black-hole stable branch by an intermediate unstable segment, which is consistent with the Van der Waals--type phase structure in the extended phase space. Figures \ref{fig:specific_heat_4panels}(c) and \ref{fig:specific_heat_4panels}(d) show the impact of the parameter $\gamma$: larger $\gamma$ suppresses the effective charge term through the factor $e^{-\gamma}$, which typically shifts the poles and smooths the variation of $C_{P}$, altering the boundaries of local stability.

A comparison between $\eta=+1$ and $\eta=-1$ reveals that the stability pattern is sector dependent. For $\eta=+1$ the coexistence of stable branches separated by a pole is more readily realized in the parameter range displayed, while for $\eta=-1$ the pole structure and the extent of stable regions can be substantially altered, reflecting a different balance among the string-cloud, charge, and pressure contributions. Overall, Fig.~\ref{fig:specific_heat_4panels} provides a compact map of local stability and complements the global phase analysis obtained from the Gibbs free energy.

In terms of entropy $S$, the black hole mass given in (\ref{b3}) can be expressed as
\begin{equation}
M=\frac{1}{2}\sqrt{\frac{S}{\pi}}\,\left(1-\alpha+\eta\,e^{-\gamma}\,\pi\,\frac{Q^2}{S}+\frac{8P}{3}\,S\right).\label{b8}
\end{equation}

In the extended phase space, the first law of black hole thermodynamics and the Smarr formula for Mod(A)Max-AdS black holes reads \cite{kubiznak2012}:
\begin{equation}
    dM=T_H\,dS+V\,dP+\Phi\,dQ,\label{b9}
\end{equation}
where
\begin{align}
&T_H=\left(\frac{\partial M}{\partial S}\right)_{P, Q, \lambda}=T,\nonumber\\
&V=\left(\frac{\partial M}{\partial P}\right)_{S, Q, \lambda}=\frac{4}{3}\sqrt{\frac{S^3}{\pi}}=\frac{4\pi}{3} r^3_h.\label{b10}\\
&\Phi=\left(\frac{\partial M}{\partial Q}\right)_{P, S, \lambda}=\pi\,\eta\,e^{-\gamma}\,\frac{Q}{S}\,\sqrt{\frac{S}{\pi}}=\eta\,e^{-\gamma}\,\frac{Q}{r_h},\label{b11}
\end{align}
With these one can easily versify that
\begin{equation}
    2(T\,S-P\,V)+Q\,\Phi=M,\label{b12}
\end{equation}
which confirm coincides with the Smarr formula.

\section{Thermodynamic Criticality}

The Reissner-Nordstr\"om--AdS black hole exhibits thermodynamic criticality analogous to the liquid-gas transition of a Van der Waals fluid. Its critical behavior, characterized by specific critical temperature, pressure, and volume, was first demonstrated in \cite{KubiznakMann2012} and subsequently in \cite{OO2017}. In this section, we study thermodynamic criticality of Mod(A)Max-AdS black hole immersed in a cloud of strings analyzing how the geometric parameters ($\gamma, \alpha$) alter the critical points $(P_c,\,v_c,\,T_c)$. We also demonstrate that the critical ratio $(P_c v_c)/T_c$ satisfies the Van der Waals ideal fluid result.

The equation of state $P=P(T)$ using Eq. (\ref{b4}) can be obtained as,
\begin{equation}
    P=\frac{T}{2r_h}-\frac{(1-\alpha)}{8\pi r^2_h}+\eta\,e^{-\gamma}\,\frac{Q^2}{8\pi r^4_h},\label{c1}
\end{equation}
Performing a change of variable of specific volume $v=2 r_h$ into the Eq. (\ref{c1}) yields
\begin{equation}
P=\frac{T}{v}-\frac{\xi}{v^2}+\frac{\zeta}{v^4},\label{c2}
\end{equation}
where we set the parameters $\xi$ and $\zeta$ as,
\begin{align}
    \xi=(1-\alpha)/2\pi,\qquad \zeta=2 \eta e^{-\gamma} Q^2/\pi.\label{c3}
\end{align}

The critical point corresponds to an inflection point in the $P$-$v$ diagram, satisfying
\begin{align}
\left(\frac{\partial P}{\partial v}\right)_T = 0,\qquad 
\left(\frac{\partial^2 P}{\partial v^2}\right)_T = 0.\label{c5}
\end{align}

Solving these conditions using the given EoS in (\ref{c2}), one can obtain the critical volume, temperature, and pressure as follows:
\begin{align}
v_c &= \sqrt{\frac{6 \, \zeta}{\xi}}=\frac{2\sqrt{6\,\eta}\,e^{-\gamma/2}}{\sqrt{1-\alpha}}\,Q,\nonumber\\[2mm]
T_c &= \frac{4 \, \xi \, \sqrt{\xi}}{3 \, \sqrt{6 \, \zeta}}=\frac{e^{\gamma/2}\,(1-\alpha)^{3/2}}{3\sqrt{6 \eta}\,Q},\nonumber\\[1mm]
P_c &= \frac{\xi^2}{12 \, \zeta}=\frac{e^{\gamma}\,(1-\alpha)^2}{96\, \eta\,Q^2}.\label{c6}
\end{align}

The critical points derived above are real and physically significant only for $\eta = +1$; while for $\eta=-1$, they do not correspond to a physical phase transition.

The critical points ($v_c,\,T_c,\,P_c$) is influenced by the string cloud parameter $\alpha$ and the ModMax parameter $\gamma$. In The limit $\alpha=0$ corresponding to the absence of string cloud and $\gamma=0$, corresponding to the absence of ModMax nonlinear electrodynamics parameter, the critical points derived above reduces to the well-known RN-AdS black hole result \cite{KubiznakMann2012,OO2017}. One can easily show that the critical ratio $R_c=P_c\,v_c/T_c=3/8$, equal to the van der Waals ideal fluid result.

The critical points in the alternate
phase space are obtained by utilizing the standard definition, and we reduced the thermodynamic variables as follows:
\begin{align}
&T_r=\frac{T}{T_c}=\frac{\left[1-\alpha-\eta\,e^{-\gamma}\frac{Q^2}{r_h^2}+8\pi P r_h^2\right]}{4\pi r_h}\,\frac{3\sqrt{6 \eta}\,e^{-\gamma/2}\,Q}{(1-\alpha)^{3/2}},\label{c7}\\[2mm]
&v_r=\frac{v}{v_c}=\frac{2 r_h \sqrt{1-\alpha}}{2\sqrt{6 \eta}\,e^{-\gamma/2}\,Q},\label{c8}\\[2mm]
&P_r=\frac{P}{P_c}=\frac{96 \eta e^{-\gamma} Q^2}{(1-\alpha)^2}\,\left[\frac{T}{2r_h}-\frac{(1-\alpha)}{8\pi r^2_h}+\eta\,e^{-\gamma}\,\frac{Q^2}{8\pi r^4_h}\right].\label{c9}
\end{align}

\begin{figure*}[tbhp]
\centering
\includegraphics[scale=0.4]{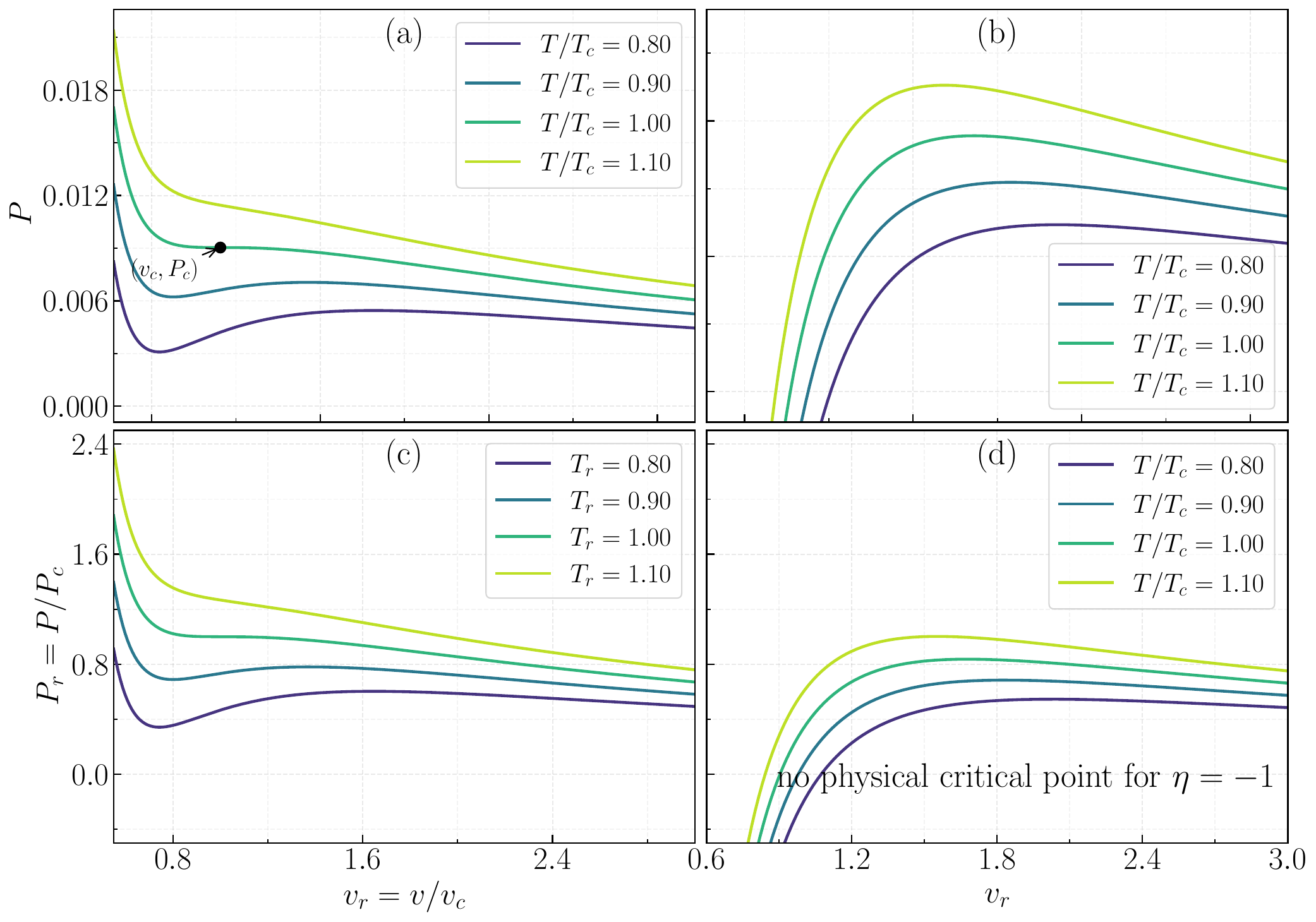}
\caption{Equation of state and criticality structure in the extended phase space for the Mod$(A)$ black hole with a cloud of strings. 
The specific volume is identified as $v=2r_h$. Panels (a) and (b) display the isotherms $P(v)$ for fixed $\alpha=0.1$, $\gamma=0.5$, and $Q=0.7$, using reduced temperatures $T/T_c=\{0.80,0.90,1.00,1.10\}$ (computed from the $\eta=+1$ critical point). Panel (a) corresponds to $\eta=+1$ and exhibits a van der Waals--like behavior with the critical point $(v_c,P_c)$ marked.
Panel (b) corresponds to $\eta=-1$, for which the isotherms do not develop a physical inflection-point criticality. Panel (c) shows the universal reduced equation of state in terms of $P_r=P/P_c$, $T_r=T/T_c$, and $v_r=v/v_c$ for $\eta=+1$, 
$P_r=\frac{8T_r}{3v_r}-\frac{2}{v_r^{2}}+\frac{1}{3v_r^{4}},
$
highlighting the mean-field critical scaling. Panel (d) illustrates the absence of a genuine reduced scaling for $\eta=-1$ (no physical critical point).}
\label{fig:eos-criticality}
\end{figure*}
Figure~\ref{fig:eos-criticality} summarizes the criticality encoded in the extended-phase-space equation of state, written in terms of the specific volume $v=2r_h$. 
For $\eta=+1$ the coefficient of the $v^{-4}$ term is positive, allowing the standard inflection-point conditions
$\left(\partial P/\partial v\right)_{T}=0$ and $\left(\partial^{2}P/\partial v^{2}\right)_{T}=0$ to be satisfied at $(v_c,T_c,P_c)$; consequently, the isotherms in Fig. \ref{fig:eos-criticality}(a) reproduce the familiar van der Waals structure, with the critical point explicitly indicated. 
The reduced variables $(P_r,T_r,v_r)$ collapse the thermodynamics into the universal mean-field form shown in Fig. \ref{fig:eos-criticality}(c), evidencing that the critical behavior is governed by the same scaling as the Reissner--Nordstr\"om--AdS class. 
In contrast, for $\eta=-1$ the sign of the $v^{-4}$ contribution changes, preventing the emergence of a physical inflection point; accordingly, Fig. \ref{fig:eos-criticality}(b) does not display the van der Waals oscillatory segment, and Fig. \ref{fig:eos-criticality}(d) emphasizes that no genuine reduced scaling can be defined because $(v_c,T_c,P_c)$ ceases to be physical in this sector.

\section{Heat Engine}

Two adiabatic and two isothermal processes combined to form the Carnot cycle are the hallmark of the most effective heat engine. The single most fundamental and critical feature of the Carnot cycle is that the heat engine efficiency is a function of reservoir temperatures \cite{Fatima2025}:
\begin{equation}
    \eta_{\rm HE}=1-\frac{T_c}{T}\label{zz1}
\end{equation}
and as a reservoir can never be at zero temperature, the efficiency cannot be one because \(T_c\) and \(T_h=T\) denote cold and hot reservoirs, respectively.

In our case at hand, we find 
\begin{equation}
    \eta_{\rm HE}=1-\frac{e^{\gamma/2}\,(1-\alpha)^{3/2}}{3\sqrt{6 \eta}\,Q}\,\frac{4\pi r_h}{1-\alpha-\eta e^{-\gamma}\,\frac{Q^2}{r^2_h}+8\pi P r^2_h}.\label{zz2}
\end{equation}
In terms of entropy $S$, we can re-write 
\begin{equation}
    \eta_{\rm HE}=1-\frac{e^{\gamma/2}\,(1-\alpha)^{3/2}}{3\sqrt{6 \eta}\,Q}\,\frac{4\sqrt{\pi S}}{1-\alpha-\eta e^{-\gamma}\,\frac{\pi Q^2}{S}+8 P S}.\label{zz3}
\end{equation}
Now, we study the behavior of the heat engine efficiency
$\eta$ as a function of the the pressure \(P\) and entropy \(S\) is illustrate.

\section{Joule-Thomson Expansion:Inversion Temperature}

In this section, we study the JT expansion of a charged AdS black hole with a string cloud in bumblebee gravity. In the extended phase space, where the black hole mass plays the role of enthalpy, the Joule-Thomson process is naturally represented by isenthalpic curves $M=\mathrm{const.}$ in the $T$--$P$ plane. For brevity, we will still refer to these as ``isenthalpic'' curves, even though, strictly speaking, they are constant-mass trajectories in the black hole parameter space. Joule--Thomson expansion in AdS black hole backgrounds has been extensively investigated for Reissner-Nordström-AdS, Kerr-AdS, Gauss-Bonnet, Lovelock, and other deformed black holes, providing a useful benchmark for our Mod(A)Max-AdS black hole setup with a string cloud (see \cite{JPMG2023,GM2024,GM2025,FA2025}).

Recall the expression for the Joule Thomson coefficient \cite{BZM1997,SQL2018,CL2020}
 \begin{equation}
  \mu_{JT}=\frac{\partial T}{\partial P}=\frac{1}{C_{P}}\left[T \left(\frac{\partial V}{\partial T}\right)_P-V\right].\label{pp1}   
 \end{equation} 
From this, we obtain the inversion temperature for zero coefficient ($\mu_J=0$) as,
\begin{equation}
    T_i=V \left(\frac{\partial T}{\partial V}\right)_P.\label{pp2} 
\end{equation}
\begin{figure*}[t]
    \centering
    \includegraphics[scale=0.35]{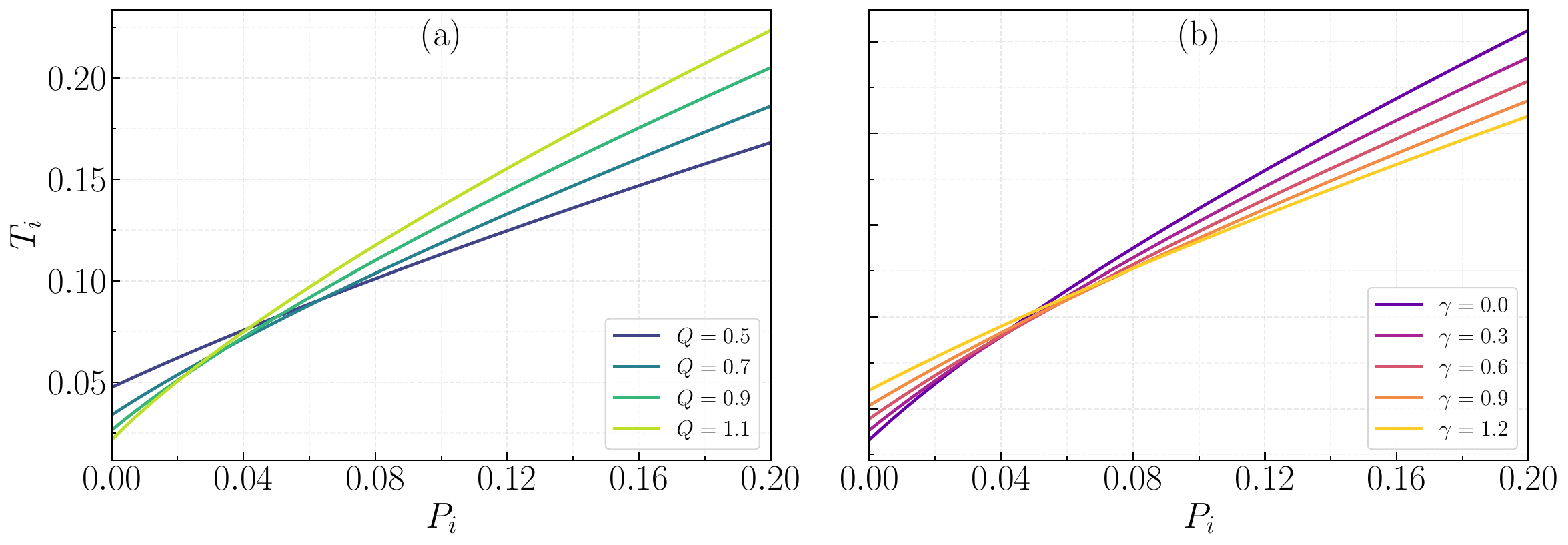}
    \caption{Joule--Thomson inversion curves $T_i$ versus $P_i$ for the Mod(A)Max--AdS black hole with a cloud of strings, obtained from the condition $\mu_{JT}=0$.
    Panel (a) shows the effect of the charge parameter $Q$ at fixed $\gamma=0.5$, while panel (b) displays the dependence on the nonlinearity parameter $\gamma$ at fixed $Q=0.7$.
    In both panels we set $\eta=+1$ and $\alpha=0.1$ (the remaining parameters are the same as those adopted throughout this section).}
    \label{fig:JT_inversion}
\end{figure*}
The Joule--Thomson (JT) expansion is characterized by the JT coefficient $\mu_{JT}=(\partial T/\partial P)_H$, whose vanishing determines the inversion curve separating heating ($\mu_{JT}<0$) from cooling ($\mu_{JT}>0$) regions.
Figure~\ref{fig:JT_inversion} displays the inversion temperature $T_i$ as a function of the inversion pressure $P_i$ for $\eta=+1$.
We observe that $T_i$ increases monotonically with $P_i$, and the family of curves is sensitive to both the charge parameter $Q$ and the nonlinearity parameter $\gamma$:
in Fig. \ref{fig:JT_inversion}(a), larger $Q$ shifts the inversion curve upward, raising the inversion temperature at a given pressure, whereas in Fig. \ref{fig:JT_inversion}(b) increasing $\gamma$ produces a systematic change in the slope and position of the inversion curve.
These behaviors indicate that the charge sector and the nonlinear electrodynamics corrections provide efficient control parameters for the JT cooling/heating boundary in the present model.
For $\eta=-1$ we do not find a physical inversion curve in the domain $P_i>0$, hence the JT inversion structure is effectively realized only in the $\eta=+1$ branch.

Now, we express the Hawking temperature obtained in Eq.~(\ref{b4}) in terms of thermodynamic volume $V=\frac{4\pi}{r} r^3_h$ as
\begin{equation}
    T=\frac{1-\alpha}{(4\pi)^{2/3} 3^{1/3}}\,V^{-1/3}
\;-\;
\frac{\eta e^{-\gamma} Q^2}{3\,V}
\;+\;
2P\left(\frac{3}{4\pi}\right)^{1/3} V^{1/3}.\label{pp4}
\end{equation}

Therefore, the inversion temperature (\ref{pp2}) simplifies to
\begin{equation}
T_i=-\frac{1-\alpha}{3(4\pi)^{2/3} 3^{1/3}}\,V^{-1/3}+\frac{\eta e^{-\gamma} Q^2}{3 V}+\frac{2P_i}{3}\left(\frac{3 V}{4\pi}\right)^{1/3} .\label{pp5}
\end{equation}
In terms of event horizon, $r_h$, we can re-write this as,
\begin{equation}
T_i=\frac{1}{4\pi r_h}\left[-\frac{1-\alpha}{3}+\frac{\eta e^{-\gamma} Q^2}{r_h^2}+\frac{8\pi P_i}{3}\,r_h^2\right].\label{pp6}
\end{equation}

At $P=P_i$, the temperature $T$ in (\ref{pp4}) simplifies to $T=T_i$ and is given by
\begin{equation}
    T_i=\frac{1-\alpha}{4\pi r_h}-\eta e^{-\gamma}\,\frac{Q^2}{4\pi r^4_h}+2 r_h P_i.\label{pp7}
\end{equation}

Subtracting Eq.~(\ref{pp6}) from (\ref{pp7}), we obtain ($r^2_h=y$)
\begin{equation}
2(1-\alpha) y-3 \eta e^{-\gamma} Q^2+8\pi P_i y^2=0.\label{pp8}
\end{equation}
The positive and real root is given by
\begin{equation}
    r_h=\frac{1}{2\sqrt{2}}\sqrt{\frac{\sqrt{(1-\alpha)^2+24 \pi \eta e^{-\gamma} Q^2\,P_i}}{\pi P_i}-\frac{1-\alpha}{\pi P_i}}.\label{pp9}
\end{equation}
Substituting $r_h$ into the Eq. (\ref{pp7}) result
\begin{align}
    T_i&=\sqrt{\frac{P_i}{2\pi}}\,\frac{1}{\left[\sqrt{(1-\alpha)^2+24 \pi \eta e^{-\gamma} Q^2\,P_i}-(1-\alpha)\right]}\,\Big[1-\alpha\nonumber\\
    &-\frac{8\pi \eta e^{-\gamma}\,Q^2\,P_i}{\left[\sqrt{(1-\alpha)^2+24 \pi \eta e^{-\gamma} Q^2\,P_i}-(1-\alpha)\right]}\nonumber\\
    &+\sqrt{(1-\alpha)^2+24 \pi \eta e^{-\gamma} Q^2\,P_i}-(1-\alpha)\Big].\label{pp10}
\end{align}
From the above expression for inversion temperature, it is clear that the string cloud parameter $\alpha$ and ModMax parameter $\gamma$ (with $\eta=+1$) influences this temperature, and thus, modifies in comparison to those for charged AdS black holes \cite{OO2017}.

At zero inversion pressure, $P_i=0$, from Eq. (\ref{pp8}) we find
\begin{equation}
    r_h=\sqrt{\frac{3 \eta e^{-\gamma}}{2(1-\alpha)}}\,Q=.\label{pp11}
\end{equation}
For real $r_h$, we must have $\eta=+1$ since $\alpha<1$. Therefore, the minimum inversion temperature from Eq. (\ref{pp7}) at zero inversion pressure is given by
\begin{equation}
    T^{\rm min}_i=\frac{(1-\alpha)^{3/2}}{6\pi \sqrt{6 \eta}\,Q}.\label{pp12}
\end{equation}
With this, one can determine and check the ratio between the minimum inversion and critical temperatures and in our case at hand, we find this ratio as,
\begin{equation}
    \frac{T^{\rm min}_i}{T_c}=\frac{1}{2}
\end{equation}
which matches with the result for charged AdS black holes \cite{OO2017}.
\begin{figure*}[tbhp]
  \centering
  \includegraphics[scale=0.4]{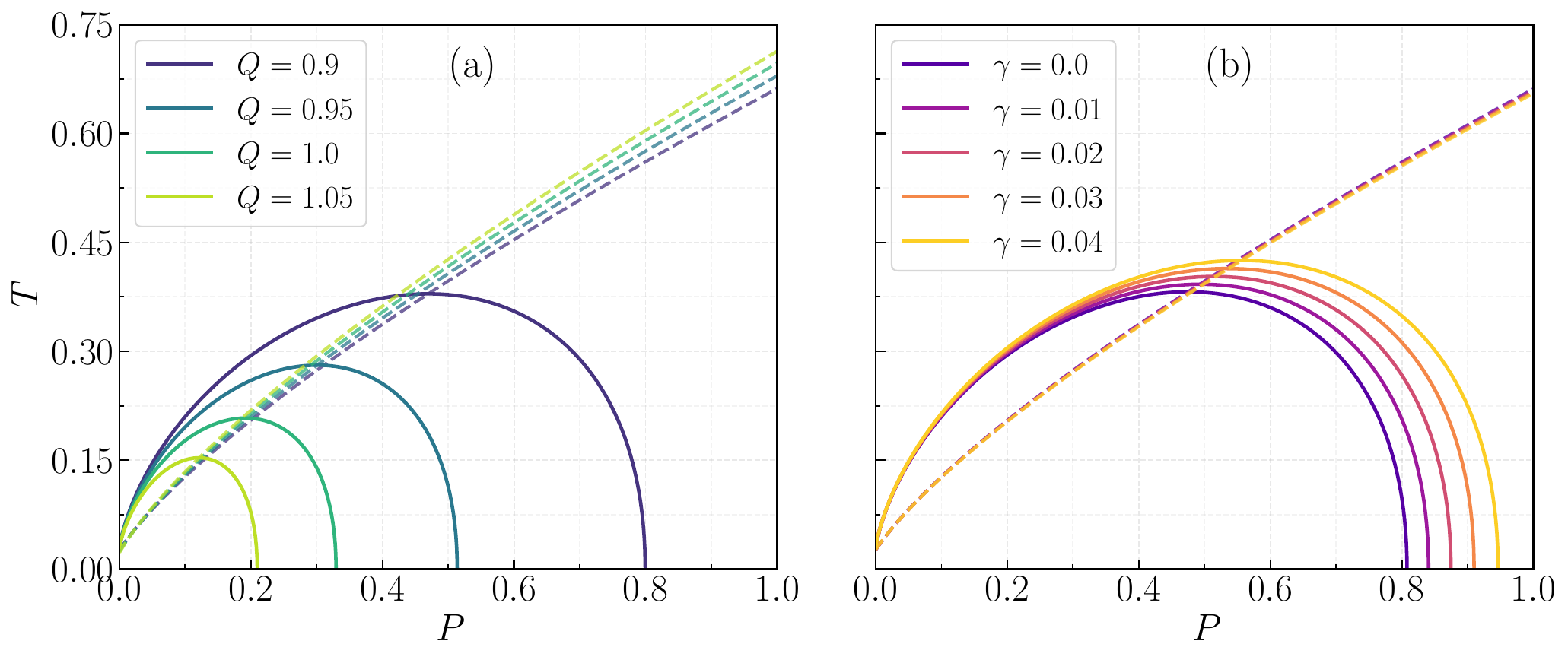}
  \caption{Joule--Thomson isenthalpic curves in the $T$--$P$ plane for fixed enthalpy $M\equiv H$ (solid curves) and the corresponding inversion curve $T_i(P_i)$ (dashed curves), computed from Eqs.~(41)--(42) using the physical root for $y=r_h^2$. (a) Isenthalpic trajectories $T(P)\big|_{M=M_0}$ for $\eta=+1$ with $(\gamma,\alpha)=(0.5,0.1)$ fixed, for several values of the charge $Q$. As $Q$ increases, the curves develop a turning point followed by a decaying/terminating branch at higher pressure, with an onset around $Q\sim Q_\ast$ (in the present scan, between $Q=0.95$ and $Q=1.0$), in qualitative agreement with the terminating isenthalpic behavior reported in related charged AdS JT analyses. (b) Isenthalpic trajectories for $\eta=+1$ with $(Q,\alpha)=(0.7,0.1)$ fixed while varying $\gamma$, showing that the nonlinear parameter shifts the location and extent of the turning/terminating branch. In both panels, intersections between solid and dashed curves identify the inversion points $(P_i,T_i)$ separating cooling/heating regions along JT expansion at fixed enthalpy.}
\label{fig:JT_isenthalpic}
\end{figure*}
\paragraph{Joule--Thomson expansion: isenthalpic trajectories and inversion curve.}
In the extended phase space, the ADM mass of the black hole plays the role of enthalpy, $H\equiv M$, so that the Joule--Thomson (JT) expansion is naturally described by \emph{isenthalpic} curves in the $T$--$P$ plane. Figure~\ref{fig:JT_isenthalpic} shows the parametric trajectories $T(P)\big|_{M=M_0}$ (solid curves), obtained by taking the horizon radius $r_h$ as the running parameter: at fixed $M_0$, one first solves the enthalpy relation for the pressure, $P=P(r_h;M_0,Q,\gamma,\eta,\alpha)$, and then evaluates the Hawking temperature $T=T(r_h;P,Q,\gamma,\eta,\alpha)$. The JT inversion curve $T_i(P_i)$ (dashed curves in Fig.~\ref{fig:JT_isenthalpic}) is computed independently from Eqs.~(\ref{pp8})--(\ref{pp9}) by selecting the physical (positive) root for $y=r_h^2$; it determines the locus $\mu_{JT}=0$ and separates the cooling/heating regions along isenthalpic expansions.

Figure \ref{fig:JT_isenthalpic}(a) is constructed for $\eta=+1$ with $(\gamma,\alpha,M)=(0.5,0.1,M_0)$ fixed, while varying the charge $Q$. A clear qualitative change emerges as $Q$ increases: for comparatively smaller $Q$ (e.g. $Q\simeq0.9$), the isenthalpic trajectory extends to higher pressures without forming a pronounced terminating branch in the plotted window, whereas for larger $Q$ the curve develops a \emph{turning point} followed by a \emph{decaying/terminating branch} at higher $P$. From the present plot, this behavior becomes prominent for $Q\gtrsim Q_\ast$ with $Q_\ast$ of order unity (in our scan, the onset is visible between $Q=0.95$ and $Q=1.0$), where the solid curve reaches a maximum temperature $T_{\max}$ at a pressure $P_{\rm turn}$ and then bends downward as $P$ increases further. Visually, for $Q\approx 1.0$ one finds a peak around $T_{\max}\sim (0.3$--$0.4)$ at $P_{\rm turn}\sim(0.3$--$0.5)$, while for $Q\approx 1.05$ the turning/termination shifts toward smaller $P$ and the downward branch becomes more pronounced. Physically, this ``decaying'' behavior indicates that, at fixed enthalpy, the admissible horizon branch ceases to exist (or leaves the physical domain) beyond a finite pressure, so that the isenthalpic curve cannot be continued after an endpoint. This feature is not an artifact of plotting: it is the characteristic terminating isenthalpic behavior highlighted in related charged AdS JT analyses (cf. the decaying isenthalpic curves emphasized in the reference EPJC example you pointed out), and therefore provides a direct and meaningful comparison point.

Figure \ref{fig:JT_isenthalpic}(b) shows the complementary dependence on the nonlinearity parameter $\gamma$ for $\eta=+1$ with $(Q,\alpha,M)=(0.7,0.1,M_0)$ fixed. In this case, the family of isenthalpic curves shifts systematically as $\gamma$ is increased: the peak region and the position of the downward branch drift, indicating that the nonlinear electrodynamic correction reshapes the mapping $(r_h)\mapsto(P,T)$ at fixed enthalpy. For the range displayed ($\gamma$ between $0$ and $\mathcal{O}(10^{-2}$--$10^{-1})$ in the plot), the turning points cluster in a comparable pressure interval, while the peak temperature changes mildly; in other words, $\gamma$ primarily controls the \emph{location} and \emph{extent} of the terminating branch rather than eliminating it. In both panels, the dashed inversion curves intersect the solid isenthalpic curves at well-defined inversion points $(P_i,T_i)$, organizing the trajectories into cooling/heating sectors: the portion of the isenthalpic curve lying on one side of the dashed locus corresponds to $\mu_{JT}>0$ (cooling), while the other side corresponds to $\mu_{JT}<0$ (heating), as required by the JT criterion.

 \section{Sparsity of Hawking radiation}
 
 In this section, we quantify the sparsity of Hawking radiation associated with our black hole solution. Although a black hole emits thermal radiation characterized by a temperature determined by its surface gravity thus resembling a classical black body the Hawking flux is inherently temporally discrete: particles are emitted as well-separated quanta rather than as a continuous stream. The concept of sparsity evaluates the ratio between the square of the thermal wavelength, $\lambda_T=2 \pi/T$, and the effective emitting area \(\mathcal{A}_{\rm eff}\), and is conventionally defined as \cite{Page1976df,Gray2016}:
 \begin{equation}
\label{defSpars}
    \psi =\frac{\mathcal{C}}{\Tilde{g} }\left(\frac{\lambda_t^2}{\mathcal{A}_{\rm eff}}\right),
\end{equation}
where \(\mathcal{C}\) is a dimensionless constant, \(\tilde g\) the spin degeneracy of the emitted quanta and \(\mathcal{A}_{\rm eff}=(27/4)\mathcal{A}_{\rm BH}=27\pi r_h^2\) is the effective area of the BH \cite{Page1976df,Gray2016}.

For a Schwarzschild black hole
\begin{equation}
\psi_{\rm Sch}=\frac{\mathcal{C}}{\tilde g}\,\frac{\lambda_T^2}{\mathcal{A}_{\rm eff}}
=\frac{\mathcal{C}}{\tilde g}\,\frac{64\pi^3}{27}\simeq\frac{\mathcal{C}}{\tilde g}\times 73.49.
\end{equation}
By invoking the definition of sparsity in Eq. (106) together with the temperature given in Eq. (40), we can elucidate the physical behavior of the sparsity parameter within the context of Mod(A)Max gravity coupled to CoS. This approach enables a clearer understanding of how the underlying parameters influence the emission characteristics. Accordingly, one obtains,
\begin{equation}\label{sspp}
\psi(r_h)=\frac{64\, \pi ^3 \,e^{2\gamma}\,r_h^4}{27 \left(Q^2 \eta-e^\gamma r_h^2 \left(1-\alpha-\Lambda  r_h^2\right)\right)^2}.\end{equation}
Equation (\ref{sspp}) reveals that the sparsity parameter is highly sensitive to the charge $Q$, the parameter $\eta $, the ModMax parameter $\gamma $, and the cosmological constant $\Lambda $. Notably, $\psi(r_h) $ diverges when the denominator approaches zero—corresponding to $T \to 0 $, i.e., the extremal limit. Conversely, for sufficiently large horizon radii where the $\Lambda  r_h^2 $ term dominates, the asymptotic behavior follows  $\psi(r_h) \propto r_h^{-4} $, indicating that very large AdS Mod(A)Max black holes become progressively less sparse.

\section{Conclusion}
In this work we performed a detailed thermal analysis of a Mod(A)Max--AdS black hole in the presence of a cloud of strings, emphasizing the interplay between the string-cloud parameter $\alpha$, the Mod(A)Max deformation $\gamma$, the charge $Q$, and the branch label $\eta=\pm1$. Treating $\Lambda$ as a pressure, we consistently adopted extended phase-space thermodynamics, with the ADM mass interpreted as enthalpy. This framework allowed us to discuss local stability, global phase preference, criticality, and JT expansion within a single, coherent setting.

A first outcome is the clear delineation of the physical parameter domain from the Hawking temperature. As shown in Fig.~\ref{fig:hawking_temperature_4panels}, the condition $T>0$ selects admissible black-hole branches and the zeros of $T$ determine the extremal limit. Increasing the charge $Q$ tends to enlarge the low-$r_h$ region where the temperature becomes negative and thus shifts the extremal radius to larger values, while increasing $\gamma$ suppresses the effective charge term through $e^{-\gamma}$ and typically reduces the extent of the negative-temperature region. The string-cloud parameter $\alpha$ enters through $(1-\alpha)$ and therefore acts as a direct control knob on the overall temperature scale and on the position of the extremal bound. These trends provide the baseline for the stability and phase-structure discussion, since the thermodynamic response functions are only meaningful on the $T>0$ branch.

The local stability analysis, captured by the heat capacity at constant pressure in Eq.~(\ref{b7}), is summarized in Fig.~\ref{fig:specific_heat_4panels}. The sign of $C_P$ separates stable ($C_P>0$) and unstable ($C_P<0$) regions, while the divergences mark points where $(\partial T/\partial r_h)_P=0$ and signal changes of local stability. Varying $Q$ shifts the location of the poles and reshapes the stable windows, and varying $\gamma$ moves these poles by attenuating the charge contribution. Importantly, the stability patterns differ between $\eta=+1$ and $\eta=-1$, reflecting the change in sign of the effective electromagnetic sector and leading to qualitatively different branch structures.

The global phase structure follows from the Gibbs free energy in Eq.~(\ref{b6}). Figure~\ref{fig:gibbs_4panels} shows that for $\eta=+1$ the system can exhibit a swallowtail profile over an appropriate pressure range, which is the standard hallmark of a first-order transition between small and large black holes. The coexistence temperature is fixed by the intersection of the competing branches (equal $G$), and the physically realized phase is the one with minimal Gibbs free energy at fixed $(T,P)$. As pressure is increased toward the critical regime, the swallowtail shrinks and disappears, indicating the end of the first-order line at a critical point. By contrast, for $\eta=-1$ the swallowtail structure is absent or strongly suppressed within the explored parameter range, anticipating the lack of a physical van der Waals criticality in that sector.

This expectation is confirmed analytically and numerically in the criticality analysis. Starting from the equation of state in Eq.~(\ref{c1}) (or Eq.~(\ref{c2}) in terms of $v=2r_h$), we obtained the closed-form critical quantities $(v_c,T_c,P_c)$ in Eq.~(\ref{c6}). The critical point exists as a real, physical inflection point only for $\eta=+1$, because the coefficient of the $v^{-4}$ term must be positive to satisfy the standard inflection-point conditions. The string cloud and Mod(A)Max parameters do not change the universal mean-field ratio $P_c v_c/T_c=3/8$, but they do shift the location of the critical point: $\alpha$ enters via $(1-\alpha)$, while $\gamma$ rescales the effective charge sector through $e^{-\gamma}$. The resulting van der Waals isotherms and the reduced universal equation of state are explicitly illustrated in Fig.~\ref{fig:eos-criticality}(a,c), whereas Fig.~\ref{fig:eos-criticality}(b,d) emphasizes the absence of a physical critical point for $\eta=-1$.

We then addressed the JT expansion, where the enthalpy interpretation $H\equiv M$ makes the constant-mass curves the natural isenthalpic trajectories in the $T$--$P$ plane. The inversion curve, defined by $\mu_{JT}=0$, was obtained analytically from Eqs.~(\ref{pp8})--(\ref{pp10}). Its dependence on $Q$ and $\gamma$ is displayed in Fig.~\ref{fig:JT_inversion}: larger $Q$ typically raises the inversion temperature at fixed $P_i$, while $\gamma$ systematically shifts the curve by modulating the effective charge contribution. Moreover, the combined visualization of isenthalpic trajectories (solid) and inversion curve (dashed) in Fig.~\ref{fig:JT_isenthalpic} makes the heating/cooling boundary transparent: intersections provide the inversion points $(P_i,T_i)$ for each enthalpy. A particularly notable feature in our model is the emergence, at sufficiently large $Q$, of turning/terminating isenthalpic branches that decay at higher pressure, a behavior that mirrors the decaying isenthalpic profiles emphasized in related AdS JT analyses and provides a concrete point of comparison for the Mod(A)Max$+$cloud-of-strings scenario. Consistently with the inversion-curve analysis, we do not find a physically relevant inversion structure for $\eta=-1$ in the positive-pressure domain considered.

Finally, we quantified the sparsity of Hawking radiation. Using Eq.~(\ref{sspp}), we showed that $\psi(r_h)$ is extremely sensitive to $(Q,\eta,\gamma,\Lambda)$ and diverges in the extremal limit where $T\to0$, reflecting the well-known fact that Hawking emission becomes highly intermittent near extremality. In the opposite regime of large horizon radius, the AdS term dominates and the sparsity decreases, indicating that sufficiently large AdS black holes radiate in a comparatively less sparse manner.

Overall, our results demonstrate that the cloud of strings and the Mod(A)Max deformation act as efficient control parameters for (i) the existence of physical branches and extremality, (ii) the local-stability windows, (iii) the presence of first-order transitions and criticality in the $\eta=+1$ sector, and (iv) the JT cooling/heating boundary and the occurrence of terminating isenthalpic behavior. Natural extensions of this analysis include incorporating additional conserved charges or rotation, exploring alternative thermodynamic ensembles, and examining whether the terminating isenthalpic branches admit an interpretation in terms of competing phases or constraints on admissible horizon branches in deformed electrodynamic sectors.

\section*{Acknowledgments}

F.A. acknowledges the Inter University Centre for Astronomy and Astrophysics (IUCAA), Pune, India for granting visiting associateship. E. O. Silva acknowledges the support from Conselho Nacional de Desenvolvimento Cient\'{i}fico e Tecnol\'{o}gico (CNPq) (grants 306308/2022-3), Funda\c c\~ao de Amparo \`{a} Pesquisa e ao Desenvolvimento Cient\'{i}fico e Tecnol\'{o}gico do Maranh\~ao (FAPEMA) (grants UNIVERSAL-06395/22), and Coordena\c c\~ao de Aperfei\c coamento de Pessoal de N\'{i}vel Superior (CAPES) - Brazil (Code 001).

\end{document}